\documentclass[aps,prf,superscriptaddress]{revtex4-2}
\usepackage{physics}
\usepackage{multirow}
\usepackage{placeins}
\usepackage{amsmath,amsfonts,amssymb,stackengine,graphicx}
\usepackage{mathtools}
\usepackage{commath}
\usepackage{natbib}
\usepackage{subcaption}
\usepackage{float}
\usepackage{tikz}
\usepackage{pgfplots}
\usepackage{newtxtext}
\usepackage{newtxmath}
\usepackage{hyperref}
\usepackage{cleveref}

\newcommand{\beq}{\begin{equation}}
\newcommand{\eeq}{\end{equation}}

\newcommand{\hx}{{\hat{\boldsymbol x}}}
\newcommand{\hy}{{\hat{\boldsymbol y}}}
\newcommand{\hz}{{\hat{\boldsymbol z}}}
\newcommand{\bu}{{\boldsymbol u}}
\newcommand{\Rey}{\text{\emph{Re}}}
\newcommand{\Pran}{\text{\emph{Pr}}}

\newcommand{\ee}{\mathrm{e}}
\newcommand{\ii}{\mathrm{i}}

\newcommand{\oG}{G}

\newcommand{\A}{{\mathbb A}}

\pgfplotsset{compat=1.18}

\begin{document}

\title{Stable stratification enhances transient growth in streaky shear flows}

\author{Will Oxley}
%\email{woo21@cam.ac.uk}
\affiliation{Department of Applied Mathematics and Theoretical Physics, University of Cambridge, Cambridge, CB3 0WA, United Kingdom}

\author{Rich R. Kerswell}
\email{rrk26@cam.ac.uk}
\affiliation{Department of Applied Mathematics and Theoretical Physics, University of Cambridge, Cambridge, CB3 0WA, United Kingdom}

%
% Abstract
%
\begin{abstract}

Recent work has found that the  well-known `lift-up' mechanism is not important for, and may even inhibit, the transient growth possible on streaky wall-bounded shear flows which is believed an important process in the near-wall cycle for turbulent flows. Moreover, artificially removing the wall-normal velocity has been found to unleash 3 orders of magnitude more perturbation energy growth in an unbounded streaky flow model. Motivated by this, we examine the effect of introducing stable stratification which naturally suppresses wall-normal velocities (the `vertical' shear case) and find it permits the  hugely enhanced linear energy growth predicted by simply removing the  wall-normal velocity. Alternatively, imposing stable stratification such that the spanwise velocities are suppressed (`horizontal shear') not surprisingly inhibits transient growth by weakening the active `push over' mechanism.  A formula for the critical stratification strength to completely suppress the preferred growth mechanism is determined which proves a useful predictor for what is seen in the full numerical solutions of the model. Implications for a stratified near-wall cycle are briefly discussed.

\end{abstract}

\maketitle
%
% I Introduction
%
\section{Introduction} \label{C5intro}

It is generally accepted that streamwise rolls and streaks  are important ingredients in a near-wall sustaining cycle for wall-bounded turbulence \citep[e.g.\! see the reviews][]{Robinson91, Panton01, Smits11, Jimenez12, Jimenez18}.
The generation of streaks from the rolls is commonly explained  by the (linear) transient growth `lift-up' mechanism \citep{Ell-75,Landahl80}, but how rolls are regenerated from the streaks has proven less clear due to the need to invoke nonlinearity at some point. Schoppa \& Hussain \cite{Sch-02} suggested that (linear) transient growth mechanisms on the streaks were actually more important than (linear) streak instabilities, and that it was these transiently-growing perturbations which fed back to create streaks through their nonlinear interaction. While this view has been contested \citep[e.g.][]{Hoepffner05,Cassinelli17, Jimenez18}, it is supported by recent cause-and-effect numerical experiments by \cite{Loz-21} which also found that lift up was not important in this process: see their \S6.4 and figure 24(a).  
Subsequent theoretical work \cite{MK24,OK25} has confirmed this finding with  lift-up actually found to inhibit growth (e.g. figure 7 of \cite{MK24} and figures 5 and 7 of \cite{OK25}) and offered a mechanistic explanation why (see \S 4.9 of \cite{OK25}). 

Investigating this phenomenon further, \cite{OK25} also found that by artificially removing the wall-normal velocity (a more drastic intervention than just removing the lift-up term) 3 orders of magnitude more growth could occur (see figures 8(b) and 8(c) of \cite{OK25}).  
This then  begs the question as to whether adding  wall-normal stable stratification could unlock these levels of  transient growth on streaky flows by naturally suppressing the wall-normal velocities.  The purpose of this brief note is to explore this question in a stratified version of the augmented Kelvin model developed in \cite{OK25}. The focus is on wall-normal stratification (or `vertical' shear) where growth enhancement is expected but wall-parallel stratification (or `horizontal shear') will also be studied for completeness. Beyond this specific issue, there are, of course, a multitude of reasons for studying stratified flows given their prevalence in nature (e.g. see \cite{Ahmed21}).

%Figure 7 of OK25 shows that the presence of lift-up actually hinders growth (the crossing of blue and red lines at kz ≈ 3). This effect is also present in figure 5(b) of OK25 and is consistent with the observations of Lozano-Durán et al. (2021) (e.g. fig 24(b) ) and figure 7 in Markeviciute & Kerswell (2024) + figure 8 of OK25 (compare 8(b) and (c) when lift-up removed - growth increases by 2 orders of magnitude.)

There has been plenty of work investigating the effect of stratification on transient growth mechanisms for  Kelvin's \cite{Kel-87} model of simple unbounded shear \cite{Far-93b, Bak-01, Bak-09a, Bak-09b, Sal-13}. This has been extended to non-uniform shears \cite[e.g.][]{Arratia11, Kaminski14}, flows with boundaries \cite[e.g.][]{Roy15, Parente20}, curved flows \cite[e.g.][]{Park17} and even turbulent flows \cite[e.g.][]{Zasko23, Cossu23,Variale24}. All, however, assume a 1-dimensional base flow such that the transient growth calculation is also 1-dimensional (i.e. a single wavenumber can be assumed for the perturbation in each homogeneous direction). The novelty here is to consider a richer 2-dimensional streaky base flow (i.e. it depends on two spatial directions rather than having 2 different velocity components) which  leads to a more-involved 2-dimensional transient growth calculation (e.g. see \cite{MK24} which treats an unstratified streaky flow). We follow  \cite{OK25} here who have already demonstrated that just such a base state in Kelvin's model augmented by a spanwise spatially-periodic shear (see (\ref{baseflow})\,)is readily accessible and a useful idealisation of  a streaky shear flow. The addition made here is to add a linear stable background stratification. 

In what follows below, gravity will be used to define the `vertical' while the base flow will define the coordinate system (the flow direction being $\hx$, the wall-normal direction being $\hy$ and the spanwise direction being $\hz$). The `vertical shear' case will have $\hy$ in the vertical direction - see fig. \ref{verticalshear} - while the `horizontal shear' case - see fig. \ref{horizontalshear} - will have $\hz$ in the vertical direction so $\hy$ is horizontal.

%
% II Formulation
%
\section{Problem Formulation} \label{C5form}

\subsection{Governing Equations \& Kelvin Modes}

The basic streaky flow model  is an unbounded constant shear with a periodic spanwise shear:
\begin{equation}
{\boldsymbol U}_B = U_B(y,z)\hx = [y + \beta \cos{(k_zz)}] \hx,
\label{baseflow}
\end{equation}
as studied in \cite{OK25} together with now a background stable linear stratification. The base flow has been made non-dimensional through the (main) $y$-shear rate and the initial wavelength in the $y$ direction, and $\beta$ is the dimensionless streak strength, while $k_z$ is the dimensionless wavenumber of the streaks. The dimensionless linearised Navier-Stokes equations for the perturbation velocity $\bu = (u,v,w)$, the pressure $p$ and the density $\rho$, under the Boussinesq approximation, are 
\begin{align}
\frac{\partial \bu}{\partial t} 
+[y + \beta \cos{(k_zz)}]\frac{\partial \bu}{\partial x} +[v-\beta w k_z \sin{(k_zz)}]  \hx  \,=\,
& 
-{\boldsymbol \nabla} p 
-\rho [\cos{\alpha}\,\hy +\sin{\alpha}\,\hz] 
+ \frac{1}{\Rey}{\boldsymbol \nabla^{2}} \bu, \label{mom}\\
\frac{\partial \rho}{\partial t}+[y + \beta \cos{(k_zz)}]\frac{\partial \rho}{\partial x}-N^2[v \cos{\alpha}+w\sin{\alpha}]   \,= & \,\,\frac{1}{\Pran \Rey} \boldsymbol \nabla^{2}\rho, \label{den} \\
\boldsymbol \nabla \boldsymbol \cdot \boldsymbol u\,= & \, 0 \label{div}
\end{align}
where the Brunt-V\"{a}is\"{a}l\"{a} frequency 
$N:=\sqrt{-g \rho_b^{'}/\rho_0} $
($g$ is the acceleration due to gravity, $\rho_0$ is the reference background density and $\rho_b^{'}$ is the gradient of the background density field in the opposite direction to gravity) is assumed spatially constant, $\Rey$ is the Reynolds number and $\Pran$ is the Prandtl number. The angle $\alpha$ is the inclination of the cross-shear direction $\hy$ to gravity: $\alpha=0$ is the `vertical shear' case and  $\alpha=\tfrac{1}{2}\pi$ the `horizontal shear' case. This system can be reduced to evolution equations for the wall-normal velocity, $v:=\hy \cdot {\boldsymbol u}$, and the wall-normal vorticity, $\eta:= \hy \!\cdot\! \boldsymbol \nabla \times {\boldsymbol  u}=\partial u / \partial z - \partial w/\partial x$, by taking $\hy \! \cdot \! \boldsymbol \nabla \times \boldsymbol \nabla \times$ and $\hy \!\cdot\! \boldsymbol \nabla \times$ of (\ref{mom}) respectively giving 
\begin{align}
\left[\frac{\partial}{\partial t} + [y + \beta \cos{(k_z z)}] \frac{\partial}{\partial x}  - \frac{1}{\Rey}\Delta\right] \Delta v+  2\beta k_z \sin{(k_z z)} \left[\frac{\partial^2 w}{\partial x \partial y} - \frac{\partial^2 v}{\partial x \partial z} \right]
& 
- \beta k_z^2 \cos{(k_z z)}\frac{\partial v}{\partial x}  -\sin{\alpha}\frac{\partial^2 \rho}{\partial y \partial z} \nonumber \\ 
& +\cos{\alpha}\left[\frac{\partial^2 \rho}{\partial x^2}+\frac{\partial^2 \rho}{\partial z^2}\right] =0,  \label{v}\\
\left[\frac{\partial}{\partial t} + [y + \beta \cos{(k_z z)}] \frac{\partial}{\partial x}  - \frac{1}{\Rey}\Delta\right] \eta + \frac{\partial v}{\partial z} + \beta k_z \sin{(k_z z)}\frac{\partial v}{\partial y}- & \beta w k_z^2 \cos{(k_z z)} - \sin{\alpha}\frac{\partial \rho}{\partial x}=0 \label{eta}
\end{align}
where $\Delta := \nabla^2$ is the Laplacian operator. The awkwardness of the $y$-dependent advection term can be removed by the well-known trick \cite{Kel-87} of using a time-dependent cross-shear wavenumber, while the $z$-dependence in the equations is handled straightforwardly through using a sum of Kelvin modes:
\begin{equation}
[u,v,w,p,\eta,\rho](x,y,z,t) = \sum_{m=-\infty}^\infty [\hat{u}_m,\hat{v}_m,\hat{w}_m,\hat{p}_m,\hat{\eta}_m,\rho_m](t)\ee^{\ii \left[k_xx+(1-k_xt)y+mk_z z\right]}.
\label{KWexpansion}
\end{equation}
Given $\hat{v}_m$ and $\hat{\eta}_m$, the other modal velocity components can be recovered via
\begin{equation}
\hat{u}_m = \frac{-k_x(1-k_xt)\hat{v}_m - \ii m k_z\hat{\eta}_m}{k_x^2+m^2k_z^2} \quad \text{and} \quad \hat{w}_m = \frac{-m k_z(1-k_xt)\hat{v}_m + \ii k_x\hat{\eta}_m}{k_x^2+ m^2 k_z^2}. \label{C5uweq}
\end{equation}
Inserting the Kelvin mode expansion (\ref{KWexpansion}) into equations (\ref{den}, (\ref{v}) and (\ref{eta}) leads to
%
% v
%
\begin{align}
\dot{\hat{v}}_m \hspace{0.1em} & = \hspace{0.1em} \left[\frac{2k_x(1-k_xt)}{k_m^2}- \frac{k_m^2}{\Rey}\right]\hat{v}_m 
\nonumber\\
&\hspace{2em} + \frac{\ii \beta k_x}{2k_m^2}\left[ 2(m+1)\frac{k_z^2k_{m+1}^2}{h_{m+1}^2} - k_z^2 - k_{m+1}^2 \right]\hat{v}_{m+1} -\frac{\ii \beta k_x}{2k_m^2}\left[ 2(m-1)\frac{k_z^2k_{m-1}^2}{h_{m-1}^2} + k_z^2 + k_{m-1}^2 \right]\hat{v}_{m-1} \nonumber\\ 
& \hspace{5em}+\hspace{0.1em} \frac{\beta k_x^2 k_z (1-k_xt)}{k_m^2h_{m+1}^2}\hat{\eta}_{m+1} -\frac{\beta k_x^2 k_z (1-k_xt)}{k_m^2h_{m-1}^2}\hat{\eta}_{m-1} \hspace{0.1em} +\hspace{0.1em}\frac{mk_z(1-k_x t)}{k_m^2}\hat{\rho}_m \sin{\alpha} \hspace{0.1em} -\hspace{0.1em}\frac{h_m^2}{k_m^2}\hat{\rho}_m \cos{\alpha}, \label{C5dv}\\
& \nonumber \\
%
% eta
%
 \dot{\hat{\eta}}_m \hspace{0.1em} &= \hspace{0.1em}- \hspace{0.1em}\frac{k_m^2}{\Rey}\hat{\eta}_m \hspace{0.1em} - \hspace{0.1em} \frac{\ii k_x\beta}{2}\left[1-\frac{k_z^2}{h_{m+1}^2} \right]\hat{\eta}_{m+1}\hspace{0.1em} - \hspace{0.1em}\frac{\ii k_x\beta}{2}\left[1-\frac{k_z^2}{h_{m-1}^2} \right]\hat{\eta}_{m-1} \hspace{0.1em}-  \hspace{0.1em}\ii m k_z\hat{v}_m\nonumber \\
&\hspace{4em} - \hspace{0.1em}\frac{\beta k_z(1-k_xt)}{2}\left[\frac{(m+1)k_z^2}{h_{m+1}^2}-1\right]\hat{v}_{m+1}  \hspace{0.1em}-\hspace{0.1em} \frac{\beta k_z (1-k_xt)}{2}\left[\frac{(m-1)k_z^2}{h_{m-1}^2}+1\right]\hat{v}_{m-1} + \ii k_x \hat{\rho}_m \sin{\alpha}, \label{C5deta}\\
& \nonumber \\
%
% rho
%
\dot{\hat{\rho}}_m \hspace{0.5em} &= - \hspace{0.1em} \frac{k_m^2}{\Pran \Rey}\hat{\rho}_m \hspace{0.1em} -\hspace{0.1em} \frac{N^2 m k_z (1-k_x t)}{h_m^2}\hat{v}_m \sin{\alpha}\hspace{0.1em} + \hspace{0.1em} N^2\hat{v}_m \cos{\alpha}
+\hspace{0.1em}  \frac{\ii N^2 k_x}{h_m^2}\hat{\eta}_m \sin{\alpha} \hspace{0.1em} - \hspace{0.1em} \frac{\ii \beta k_x}{2}\hat{\rho}_{m-1}\hspace{0.1em} - \hspace{0.1em} \frac{\ii \beta k_x}{2}\hat{\rho}_{m+1}. \label{C5drho}\\
& \nonumber
\end{align}
where $k_m^2:= k_x^2+(1-k_xt)^2+m^2 k_z^2\,\,$ and $\,h_m^2:=k_x^2+m^2 k_z^2$. This set of equations for sufficiently large $M$ (so the solution becomes insensitive to $M$) is referred to as the {\bf `full'} system hereafter.

\subsection{Optimal Gain, Parameter Choices \& Numerical Methods}

The volume-averaged kinetic energy and potential energy (e.g. \citep{Holliday81}) contributions are
\begin{align}
KE(t) &\hspace{0.5em}:= \hspace{0.5em} \frac{1}{V_\varOmega}\int_\varOmega \frac{1}{2} |{\boldsymbol u({\bf x},t})|^2 \,d\varOmega
\hspace{0.5em} = \hspace{0.5em} \frac{1}{2} \sum_{m=-M}^M \frac{1}{h_m^2} \left[k^2_m\abs{\hat{v}_m}^2 +\abs{\hat{\eta}_m}^2\right], \label{KE} \\
PE(t) &\hspace{0.5em}:= \hspace{0.5em}\frac{1}{V_\varOmega}\int_\varOmega\, \frac{1}{2N^2}|{\rho({\bf x},t})|^2 \,d\varOmega
\hspace{0.5em}=\hspace{0.5em} \frac{1}{2} \sum_{m=-M}^M \frac{\abs{\hat{\rho}_m}^2}{N^2}
\label{PE}
\end{align}
where $\varOmega:=[0,2\pi/k_x] \times [0,2 \pi] \times [0,2\pi/k_z]$ and $V_\varOmega$ is the volume of $\varOmega$.  The `optimal gain' is defined in terms of the  growth of the total energy $E(t):=KE(t)+PE(t)$ so that the optimal gain is the largest value of $E(T)/E(0)$ over all possible initial conditions:
\begin{equation}
\oG(T,k_x,k_z,\beta,N,\Rey,\Pran) := \max_{{\bu}(0),\rho(0)}\frac{E(T,k_x,k_z,\beta,N,\Rey,\Pran)}{E(0,k_x,k_z,\beta,N,\Rey,\Pran)}.
\label{C5optgain}
\end{equation}
The `global optimal gain' is the result of optimising this quantity over the two wavenumbers.
%\begin{equation}
%\goG(T,\beta,N,\Rey,\Pran) := \max_{\{k_x,k_z\}}\oG(T,k_x,k_z,\beta,N,\Rey,\Pran),
%\label{C5optgainglobal}
%\end{equation}
%
Choices of $T=5$, $\beta=1$, $\Rey=200$ are made throughout appropriate for the buffer or lower log layer as in \cite{OK25}. A fixed value of $\Pran=7$ is also taken so that our full focus here is in varying $N$, the strength of the stratification. After choosing an appropriate truncation value $M$, initial conditions are integrated forward in time using the modal evolution equations (\ref{C5dv}) - (\ref{C5drho}) using MATLAB's ode45. A constrained optimisation technique is used to find the optimal gain and associated optimal initial conditions \cite{OK25} for a given wavenumber pair $(k_x,k_z)$. 
% The overall optimal gain is found by using a golden section search method.

%--------------------
%
%  III Vertical shear
%
%--------------------

%
% fig 1
%
\begin{figure}
    \centering
    \begin{tikzpicture}
        % Include the PNG image
        \node[anchor=south west, inner sep=0] (img) at (0,0) {\includegraphics[width=7cm]{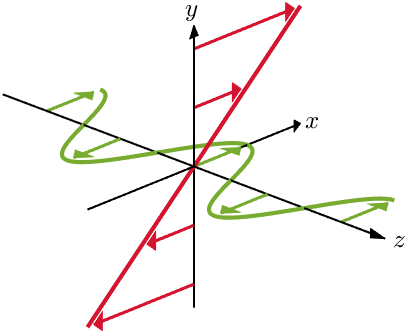}};

        % Overlay extra elements
        \begin{scope}[x={(img.south east)}, y={(img.north west)}] % Normalized coordinates (0 to 1)

            % Moved to top-left, thicker, blue arrow
            \draw[line width=2pt, ->, >=stealth, blue!70!black] (0.15, 0.95) -- (0.15, 0.74);
            \node[anchor=west, blue!70!black] at (0.18, 0.88) {\large {$\boldsymbol{g}$}};
            
        \end{scope}
    \end{tikzpicture}

\caption{The vertical unbounded constant shear is shown in red, extended by horizontal (spanwise) streaks in green, with the gravity vector indicated in blue.}
\label{verticalshear}
\end{figure}

\section{Vertical Shear} \label{vshear}

The vertical shear  situation is illustrated in fig. \ref{verticalshear} where stable stratification opposes `vertical' motions $v$. This is the interesting case as the variable $v$ has a stabilising influence on the streaky flow growth mechanism so suppressing $v$ should enhance growth. The mechanism for transient growth in unstratified streaky flow was successfully captured  by a minimal 2-variable system (equations (4.14)-(4.15) in \cite{OK25}), so the starting point here is to investigate a stratified version of this  model. Reducing the truncation to just $M=1$ and assuming sinuous symmetry so
\beq
\hat{v}_0=0, \quad \hat{v}_1=-\hat{v}_{-1}, \quad \hat{\eta}_1=\hat{\eta}_{-1}, \quad \hat{\rho}_0=0, \quad\hat{\rho}_1=-\hat{\rho}_{-1}
\eeq
(justified by the form of the initial optimal conditions in the full numerical solutions) produces a {\bf `reduced'} system which has four variables:
\begin{align}
\frac{d\hat{v}_1}{dt} &= -\frac{k_1^2}{\Rey}\hat{v}_1 + \frac{2k_x(1-k_xt)}{k_1^2}\hat{v}_1- \frac{\beta k_z (1-k_xt)}{k_1^2}\hat{\eta}_0 - \frac{h_1^2}{k_1^2}\hat{\rho}_1,  \label{C5bdv1} \\
\frac{d\hat{\eta}_0}{dt} &= -\frac{k_0^2}{\Rey}\hat{\eta}_0 - \frac{\ii \beta k_x^3}{h_1^2}\hat{\eta}_1 +\frac{\beta k_x^2 k_z(1-k_xt)}{h_1^2}\hat{v}_1, \label{C5bdeta0} \\
\frac{d\hat{\eta}_1}{dt} &= -\frac{k_1^2}{\Rey}\hat{\eta}_1  + \frac{\ii \beta (k_z^2-k_x^2)}{2k_x}\hat{\eta}_0 - \ii  k_z \hat{v}_1, \label{C5bdeta1} \\
\frac{d\hat{\rho}_1}{dt} &= -\frac{k_1^2}{\Pran\Rey}\hat{\rho}_1 +N^2\hat{v}_1. \label{C5bdrho1} 
\end{align}
\cite{OK25} also dropped the $\hat{v}_1$ variable to get their  `minimal' system. Doing the same here, however, gives a system where the density is decoupled from the vorticity equations which then reverts trivially to the unstratified system under optimisation. This just reinforces the fact that the  vertical velocity, $\hat{v}_1$, is the key ingredient here and so we work with the 4-variable `reduced' system (\ref{C5bdv1})-(\ref{C5bdrho1}) to retain it.

%
% fig 2
%
\begin{figure}
    \centering
%   \begin{tabular}{@{}m{0.8cm}@{}m{\textwidth}@{}}
%        \raisebox{24pt}{\footnotesize \textbf{$N=1$}} &
%        \begin{subfigure}{\linewidth}
            \centering
            \includegraphics[trim={0cm 0cm 0cm 0cm}, clip,width=0.9\textwidth]{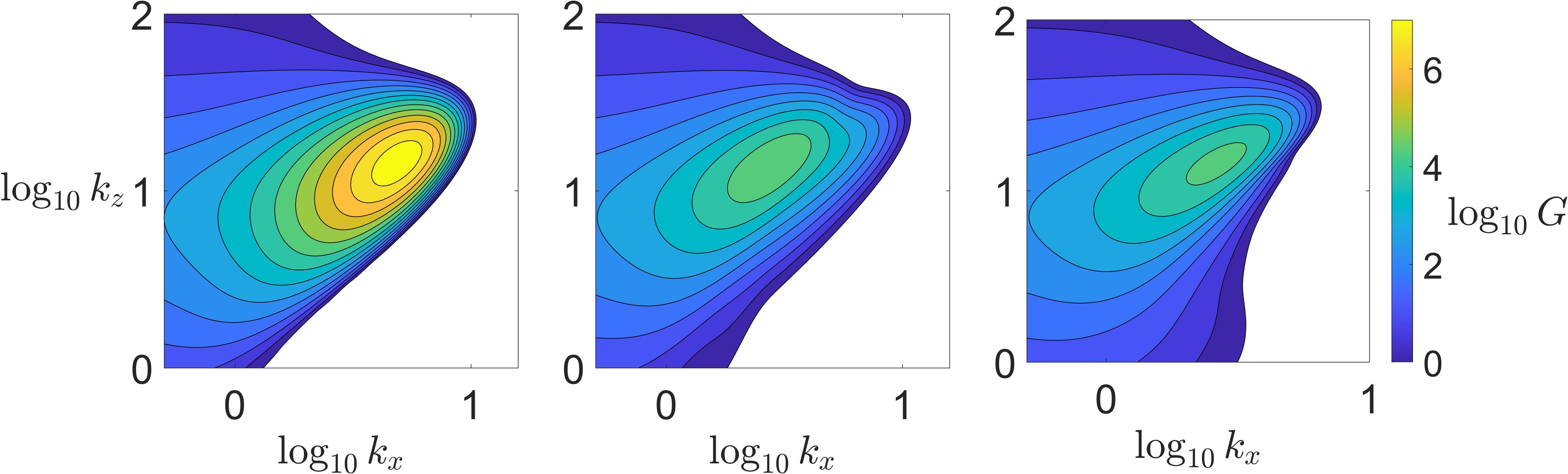}\\
%        \end{subfigure} \\
%        \raisebox{24pt}{\footnotesize\textbf{$N=20$}} &
%        \begin{subfigure}{\linewidth}
%            \centering
            \includegraphics[trim={0cm 0cm 0cm 0cm}, clip,width=0.9\textwidth]{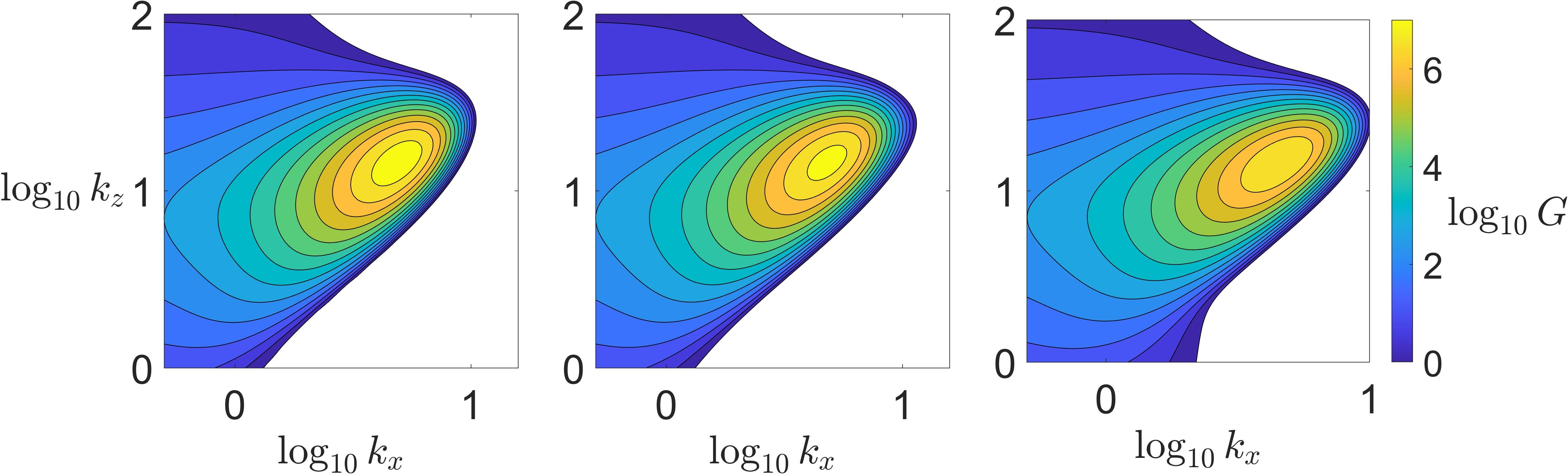}\\
%        \end{subfigure}
%    \end{tabular}
    \caption{Two sequences of contour plots of the optimal gain in wavenumber space, for parameters $T=5$, $\beta=1$, $\Rey=200$ and $\Pran=7$. The upper row has $N=1$ and the lower row $N=20$. The columns correspond to increasing complexity in the model used: the first (leftmost) is the unstratified minimal system of \cite{OK25}, the second the reduced system, and the third (rightmost) shows the full system with $M=10$. This plot shows that the reduced and full systems behave growthwise like the unstratified minimal system as the stratification gets large enough.}
    \label{vshearcomp3}
\end{figure}

\begin{figure}
\centering
\includegraphics[width=0.44\linewidth]{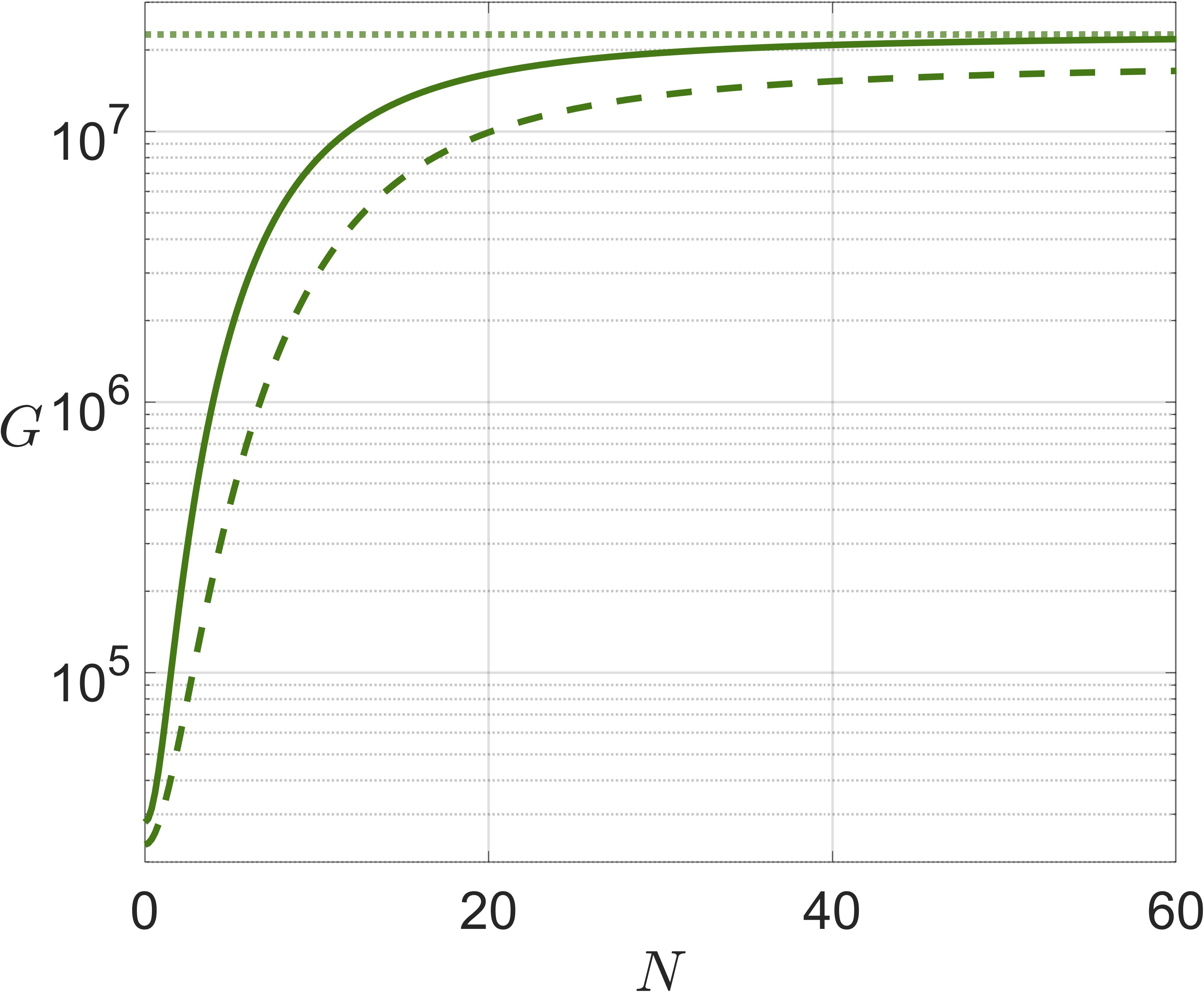}
\hspace{0.6cm}
\includegraphics[width=0.469\linewidth]{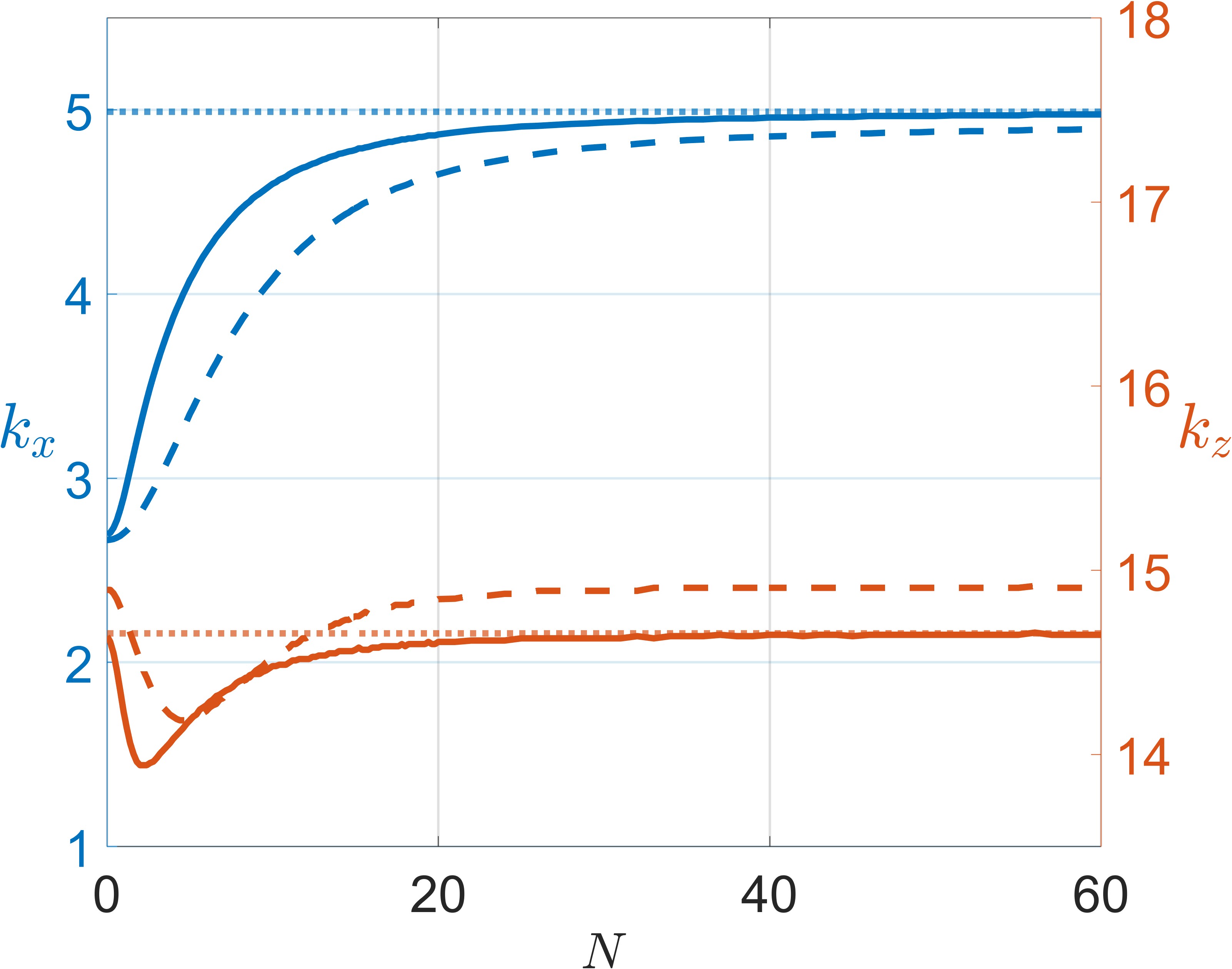}
\caption{Left: global optimal gain vs $N$ for the parameters $T=5$, $\beta=1$, $\Rey=200$ and $\Pran=7$. The solid line is for the reduced system, the dashed line is for the full system (with $M=10$) and the dotted line is for the unstratified minimal system of \cite{OK25}. Right: same as left but now for the optimising $k_x$ (blue) and $k_z$ (red). These plots demonstrate clearly that the full problem exhibits enhanced growth when stratification is increased, and that the unstratified minimal system becomes a good approximation to the full solution for large $N$. }
\label{vshearvN}
\end{figure}

Figure \ref{vshearcomp3} displays the optimal gain in wavenumber space for the reduced and full systems at moderate ($N=1$) and strong ($N=20$) stratification. Both systems clearly show enhanced  optimal gain with increasing stratification with both their energy gain maps approaching the unstratified minimal system case from \cite{OK25} (reproduced in the first column) where $\hat{v}_1$ is artificially set to 0. This indicates that the effect of stratification is exactly as expected: vertical motions which inhibit growth are themselves suppressed so growth can achieve much higher values.  To reiterate this, the growth for the full stratified system (third column of fig. \ref{vshearcomp3}) approaches the large values of the unstratified minimal system {\em not} the reduced values of the full unstratified system which are O(1000) times  smaller: see figure 8 of \cite{OK25}.
Figure \ref{vshearvN} displays the evolution of the global optimal gain and associated optimal wavenumbers for both the reduced and full systems as $N$ is increased. This confirms that both stratified systems approach the unstratified minimal system of \cite{OK25}. It also highlights the dramatic increase in global optimal growth - 3 orders of magnitude - as $N$ increases from 0 to above 20.

%
% Fig 3
%

%
% fig 4
%
\begin{figure}
    \centering
        \includegraphics[width=0.55\textwidth]{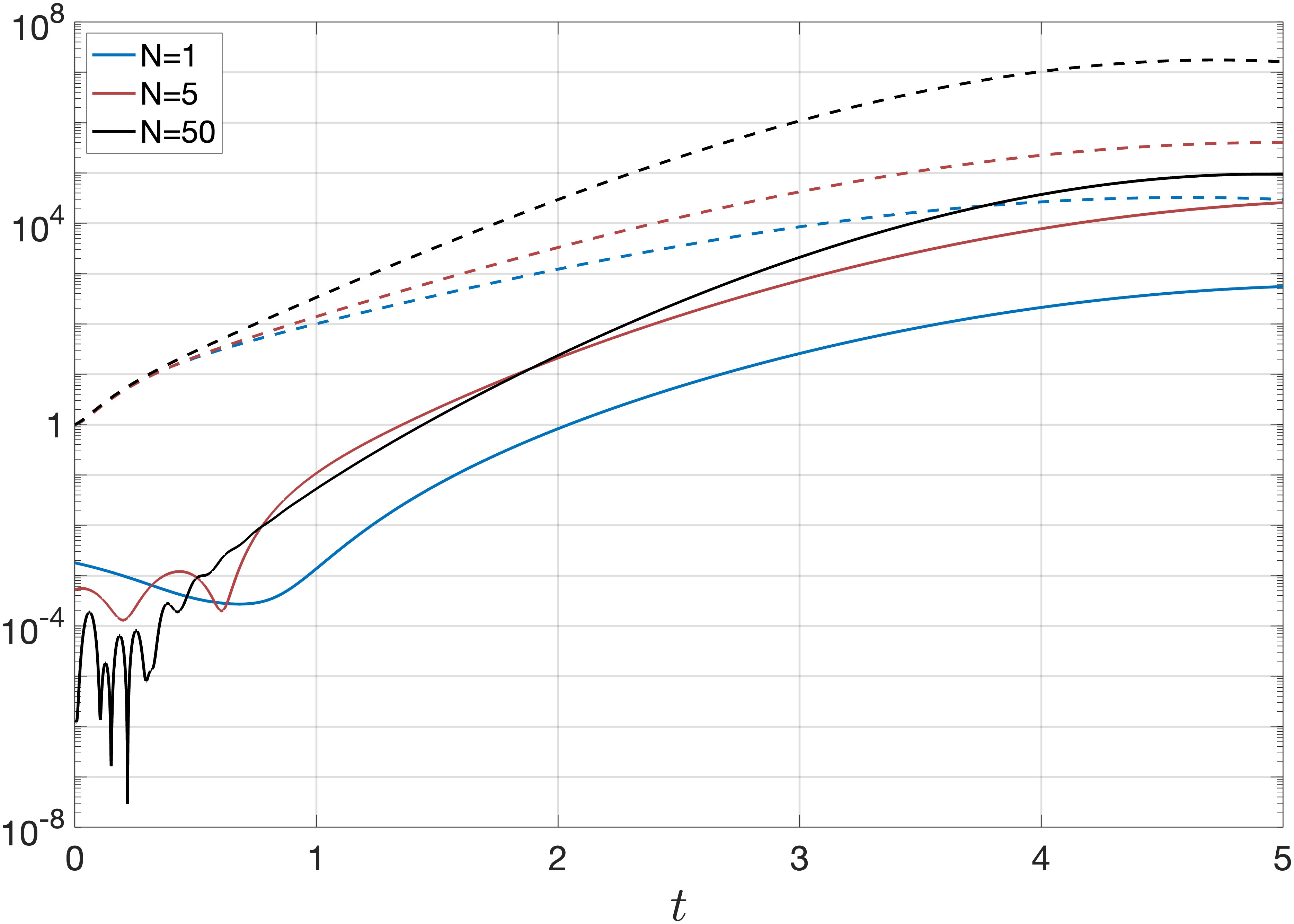}
    \hspace{0.1cm} 
% Adds space between subfigures
%        \includegraphics[width=0.36\textwidth]{Images/logEsplit_vshear_zoom.JPEG}
    \includegraphics[width=0.345\textwidth]{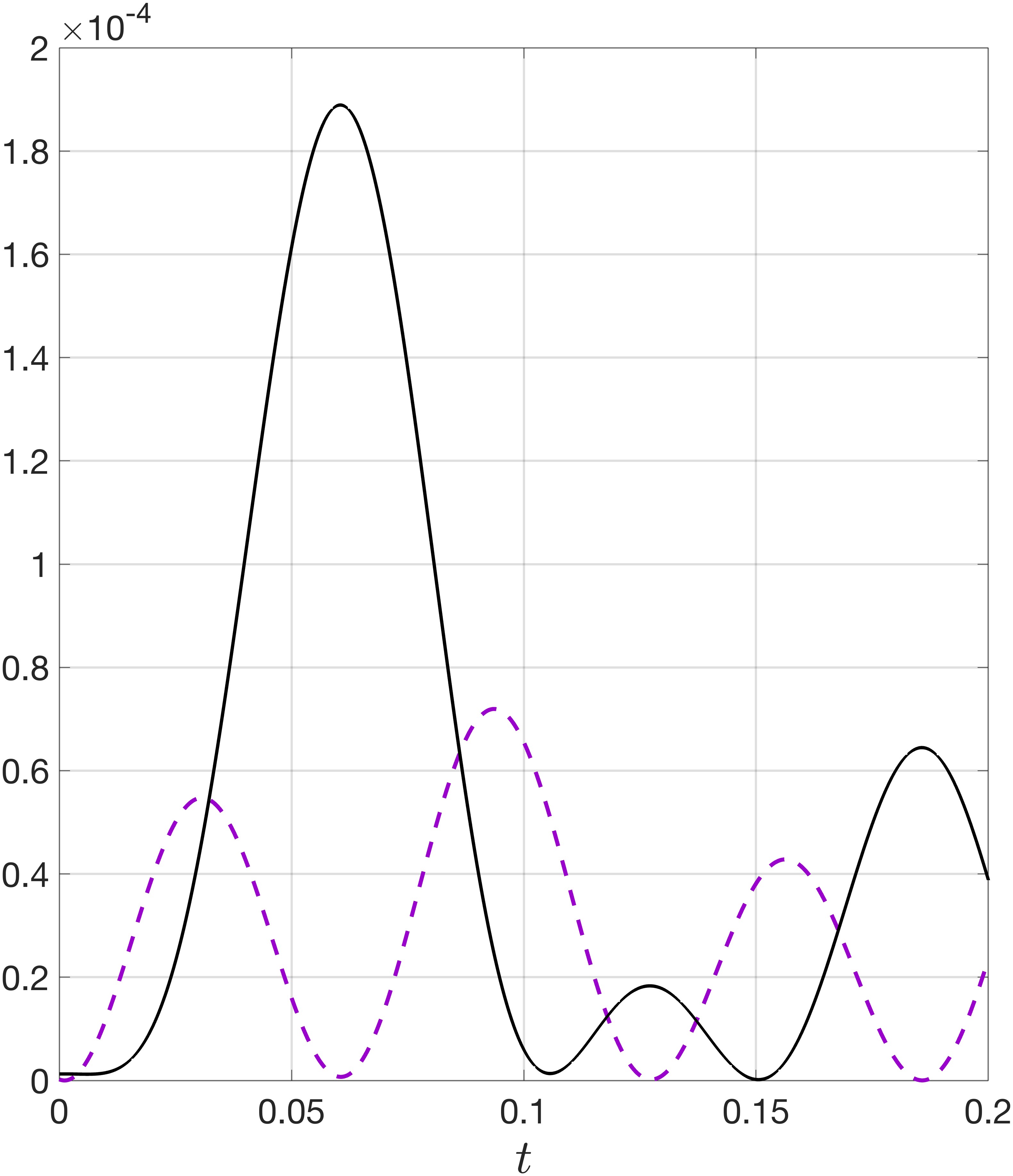}     
    \caption{Left: potential (solid lines) and kinetic (dashed lines) energies over time for the global optimal perturbation in the full system ($M=10$) for $T=5$, $\beta=1$, $\Rey=200$ and $\Pran=7$ for $N=1$ (blue), $N= 5$ (maroon) and $N=50$ (black). 
    Right: A blow-up of the potential energy (solid black again) and vertical kinetic energy (i.e. that due to $v$ alone, dashed purple) for $N=50$. This shows 
    that the vertical kinetic energy and potential energy oscillate out of phase indicating that internal waves are initially excited at $N=50$.
    }
\label{vshearenergy}
\end{figure}

%
% fig 5
%
\begin{figure}
    \centering
\includegraphics[width=0.65\textwidth]{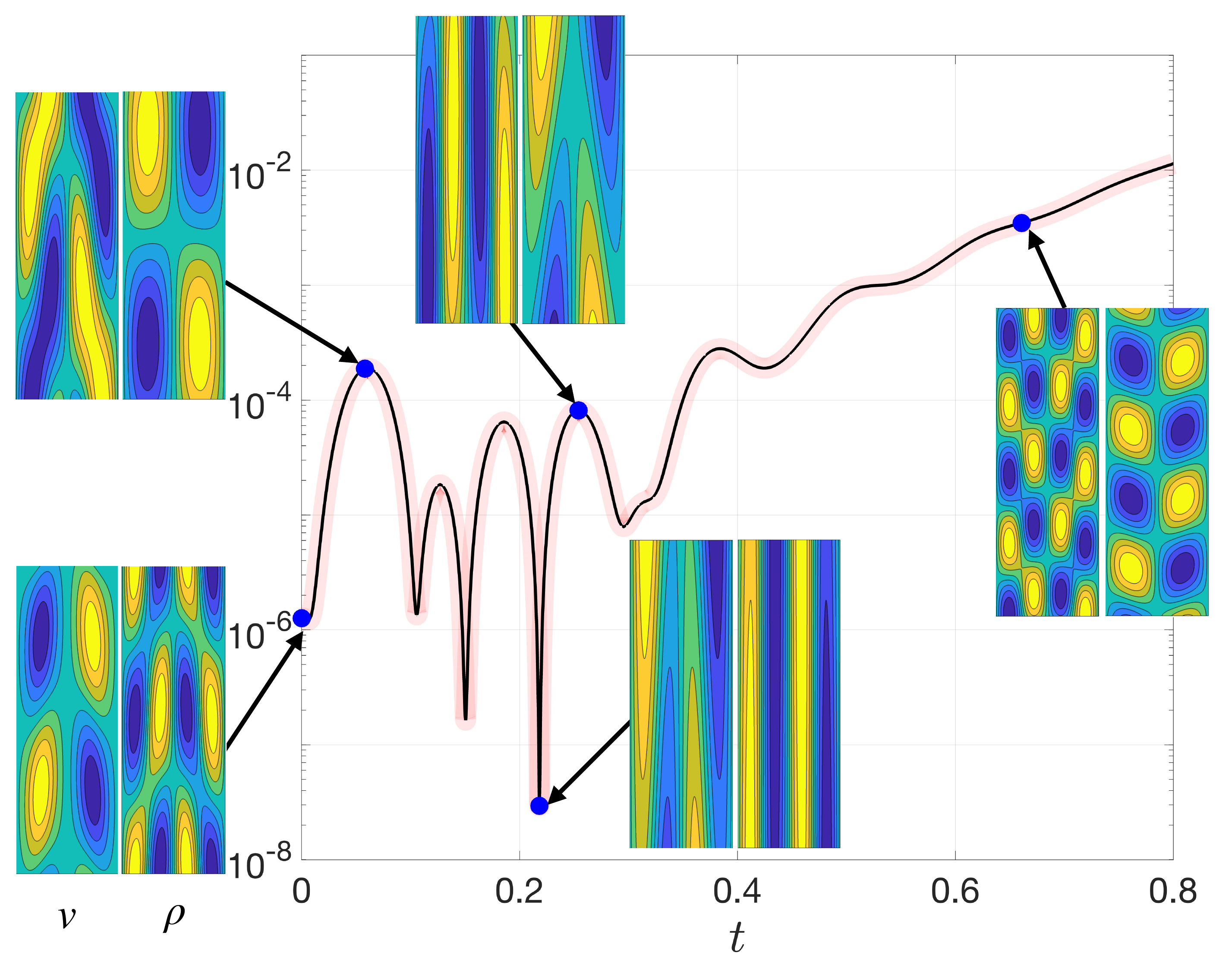}
\caption{A close-up plot of the early time potential energy evolution for $N=50$, with the solid black line showing the same data as that used for the left plot, while the transparent thick red line is used to display the data which takes $M=20$ as well as tighter integration and optimisation tolerances to confirm that the signal is real. The inset plots are of $v$ (left in each pair) and $\rho$(right) over $(z,y) \in [0,2\pi/k_z] \times [-\pi,\pi]$  at $x=0$ with the $z$ dimension magnified by a factor of 5 for clarity. The cross-shear wavenumber $k_y=1-k_x t$ vanishes at $t \approx 0.2$ here ($k_x \approx 5$ and $k_z \approx 14.7$: see figure \ref{vshearvN}(right)\,). }
\label{Focus}
\end{figure}

The breakdown of total energy into kinetic and potential contributions as a function of time for the global optimal is displayed in figure \ref{vshearenergy} using the full system ($M=10$) (the reduced system produces the same behaviour). When stratification is moderate, there appears to be two phases to the potential energy evolution with it initially decaying before growing alongside the kinetic energy: see the solid blue line for $N=1$. As the stratification increases, internal waves initially get excited  which by $N=50$ is reflected in the complicated evolution of the potential energy. Figure \ref{vshearenergy}(right) is a blow up of part of this time interval for $N=50$ and clearly shows the out-of-phase oscillations of the vertical kinetic energy and the potential energy symptomatic of internal waves. This behaviour in the kinetic energy as a whole is masked by the much larger contribution from the horizontal kinetic energy. Figure \ref{Focus} shows what the vertical velocity and perturbation density field looks like across this time period. The near-invariance with y (the y-axis in the contour plots) at about $t=0.2$ is due to the wavenumber $k_y=1-k_x t$ going through 0 ($k_x \approx 5$ for the global optimal at $N=50$ - see figure \ref{vshearvN}(right)\.) After this transition, the flow settles to sustained growth of both kinetic and potential energy (the thick transparent red line overlaying the original black line is the result of recalculating the optimal evolution with much tighter tolerances and doubling the truncation $M$ to 20).

%-------------------------------
%
% IV Horizontal Stratification
%
%-------------------------------

%
% fig 6
%
\begin{figure}
    \centering
    \begin{tikzpicture}
        % Include the PNG image
        \node[anchor=south west, inner sep=0] (img) at (0,0) {\includegraphics[width=9cm]{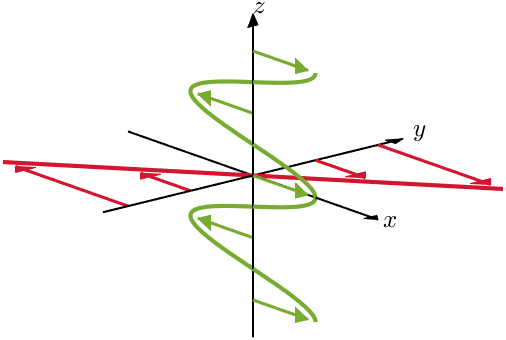}};
        % Overlay extra elements
        \begin{scope}[x={(img.south east)}, y={(img.north west)}] % Normalized coordinates (0 to 1)
            % Moved to top-left, thicker, blue arrow
            \draw[line width=2pt, ->, >=stealth, blue!70!black] (0.15, 0.95) -- (0.15, 0.74);
            \node[anchor=west, blue!70!black] at (0.18, 0.88) {{\large $\boldsymbol{g}$}};
        \end{scope}
    \end{tikzpicture}
\caption{The horizontal unbounded constant shear is shown in red, extended by vertical (spanwise) streaks in green, with the gravity vector indicated in blue.}
\label{horizontalshear}
\end{figure}

\section{Horizontal Shear} \label{hshear}

The horizontal shear  situation is illustrated in fig. \ref{horizontalshear} where stable stratification opposes `spanwise' motions, $w$, which are 
central to the transient growth process. As a result, this stratification should inhibit streaky flow transient growth and the question is then how quickly does this happen as $N$ is increased. A useful starting point to address this is the stratified version of the  minimal system found in \cite{OK25} which, contrary to the vertical shear case, is non-trivial. So, taking the drastic truncation $M=1$ and imposing the sinuous symmetry
\beq
\hat{v}_0=0,\quad \hat{v}_1=-\hat{v}_{-1}, \quad \hat{\eta}_1=\hat{\eta}_{-1}, \quad \hat{\rho}_1=\hat{\rho}_{-1}.  
\eeq
(again justified by the form of the optimal initial conditions in the full numerical solutions), produces the {\bf `reduced'} (5-variable) system:
\begin{align}
\frac{d\hat{v}_1}{dt} &= -\frac{k_1^2}{\Rey}\hat{v}_1 + \frac{2k_x(1-k_xt)}{k_1^2}\hat{v}_1- \frac{\beta k_z (1-k_xt)}{k_1^2}\hat{\eta}_0 + \frac{k_z(1-k_x t)}{k_1^2}\hat{\rho}_1,  \label{C5dv1} \\
\frac{d\hat{\eta}_0}{dt} &= -\frac{k_0^2}{\Rey}\hat{\eta}_0 - \frac{\ii \beta k_x^3}{h_1^2}\hat{\eta}_1 +\frac{\beta k_x^2 k_z(1-k_xt)}{h_1^2}\hat{v}_1+ \ii k_x \hat{\rho}_0, \label{C5deta0} \\
\frac{d\hat{\eta}_1}{dt} &= -\frac{k_1^2}{\Rey}\hat{\eta}_1  + \frac{\ii \beta (k_z^2-k_x^2)}{2k_x}\hat{\eta}_0 - \ii k_z \hat{v}_1+ \ii k_x\hat{\rho}_1, \label{C5deta1} \\
\frac{d\hat{\rho}_0}{dt} &= -\frac{k_0^2}{\Pran\Rey}\hat{\rho}_0 + \frac{\ii N^2}{k_x}\hat{\eta}_0 - \ii \beta k_x \hat{\rho}_1, \label{C5drho0} \\
\frac{d\hat{\rho}_1}{dt} &= -\frac{k_1^2}{\Pran\Rey}\hat{\rho}_1 + \frac{\ii N^2 k_x}{h_1^2}\hat{\eta}_1  -\frac{N^2 k_z (1-k_x t)}{h_1^2}\hat{v}_1- \frac{\ii \beta k_x}{2}\hat{\rho}_0. \label{C5drho1} 
\end{align}
And then finally  dropping $\hat{v}_1$ leads to  a {\bf `minimal'} (4-variable) system:
\begin{align}
\frac{d\hat{\eta}_0}{dt} &= -\frac{k_0^2}{\Rey}\hat{\eta}_0 - \frac{\ii \beta k_x^3}{h_1^2}\hat{\eta}_1 + \ii k_x \hat{\rho}_0, \label{C54vdeta0} \\
\frac{d\hat{\eta}_1}{dt} &= -\frac{k_1^2}{\Rey}\hat{\eta}_1  + \frac{\ii \beta (k_z^2-k_x^2)}{2k_x}\hat{\eta}_0 + \ii k_x\hat{\rho}_1, \label{C54vdeta1} \\
\frac{d\hat{\rho}_0}{dt} &= -\frac{k_0^2}{\Pran\Rey}\hat{\rho}_0 + \frac{\ii N^2}{k_x}\hat{\eta}_0 - \ii \beta k_x \hat{\rho}_1, \label{C54vdrho0} \\
\frac{d\hat{\rho}_1}{dt} &= -\frac{k_1^2}{\Pran\Rey}\hat{\rho}_1 + \frac{\ii N^2 k_x}{h_1^2}\hat{\eta}_1 - \frac{\ii \beta k_x}{2}\hat{\rho}_0. \label{C54vdrho1} 
\end{align}
Dropping the diffusion terms (as two are time-dependent through $k_1=\sqrt{k_x^2+(1-k_x t)^2+k_z^2}$)\,), leads simply to the eigenvalue problem 
\beq
\frac{d\boldsymbol V}{dt} = \sigma \boldsymbol V = \A \boldsymbol V, \label{C5eig0}
\eeq
where
\begin{equation}\boldsymbol V :=  
\begin{pmatrix}
\hat{\eta}_0 \\
\\
\hat{\eta}_1 \\
\\
\hat{\rho}_0 \\
\\
\hat{\rho}_1
\end{pmatrix}
\quad \text{and}  \quad \A := 
\renewcommand{\arraystretch}{2}
\begin{pmatrix}
0 & \dfrac{-\ii \beta k_x^3 }{h_1^2} & \ii k_x & 0 \\
\dfrac{\ii\beta (k_z^2-k_x^2)}{2k_x} & 0 & 0 & \ii k_x \\
\dfrac{\ii  N^2}{k_x} & 0 & 0 & -\ii  \beta k_x \\
0 & \dfrac{\ii  N^2 k_x}{h_1^2} & \dfrac{- \ii  \beta k_x}{2} & 0
\end{pmatrix}.
\label{C5eig1}
\end{equation}
%
% fig 7
%
\begin{figure}
    \centering
    \begin{subfigure}{\textwidth}
        \centering
        \includegraphics[width=\textwidth]{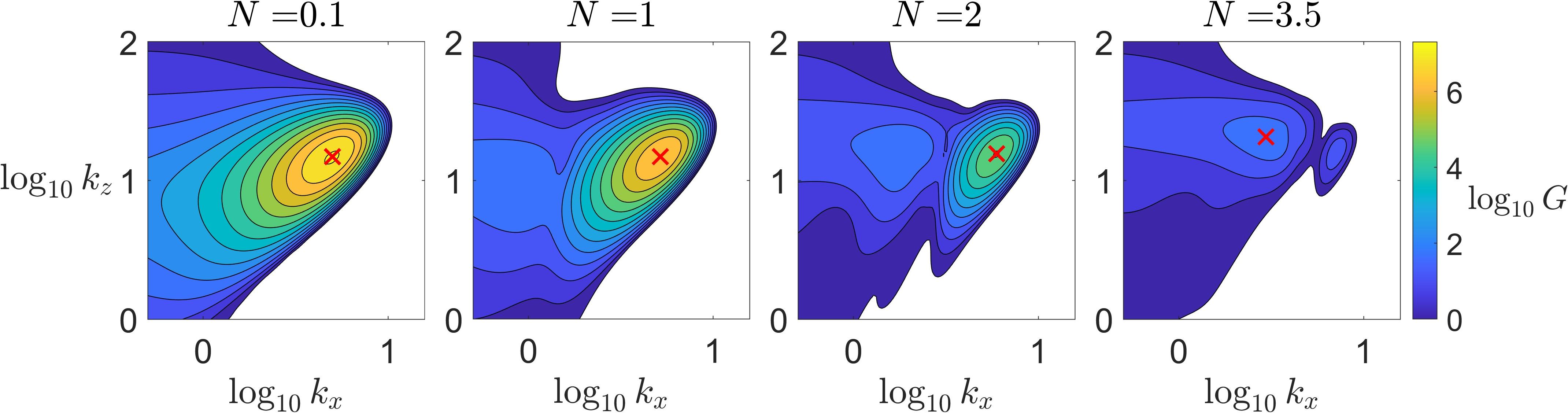}
        %\caption{}
    \end{subfigure} 
    \par\bigskip
    \begin{subfigure}{\textwidth}
        \centering
        \includegraphics[width=\textwidth]{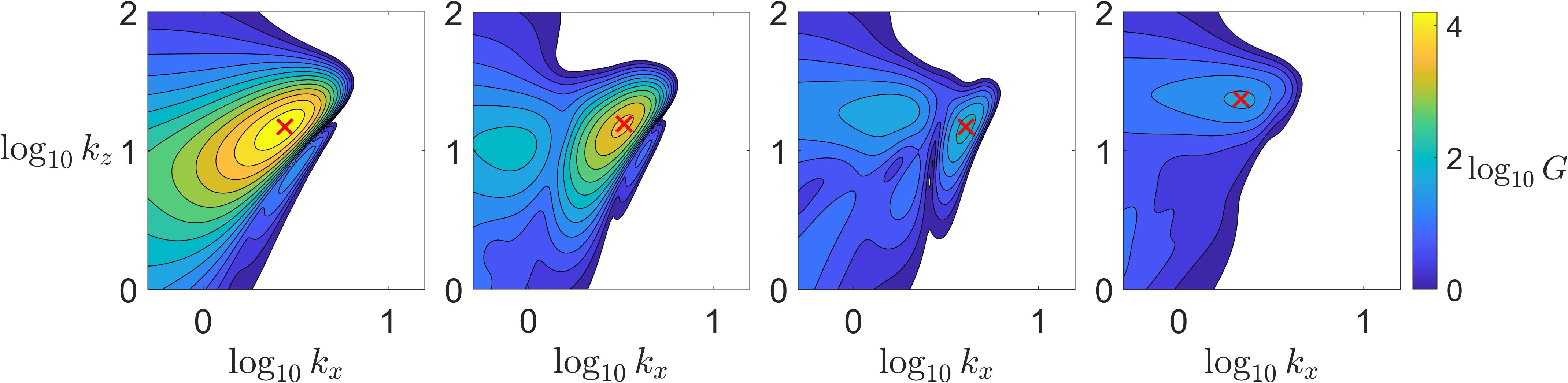}
        %\caption{}
    \end{subfigure}
    \par\bigskip
    \begin{subfigure}{\textwidth}
        \centering
        \includegraphics[width=\textwidth]{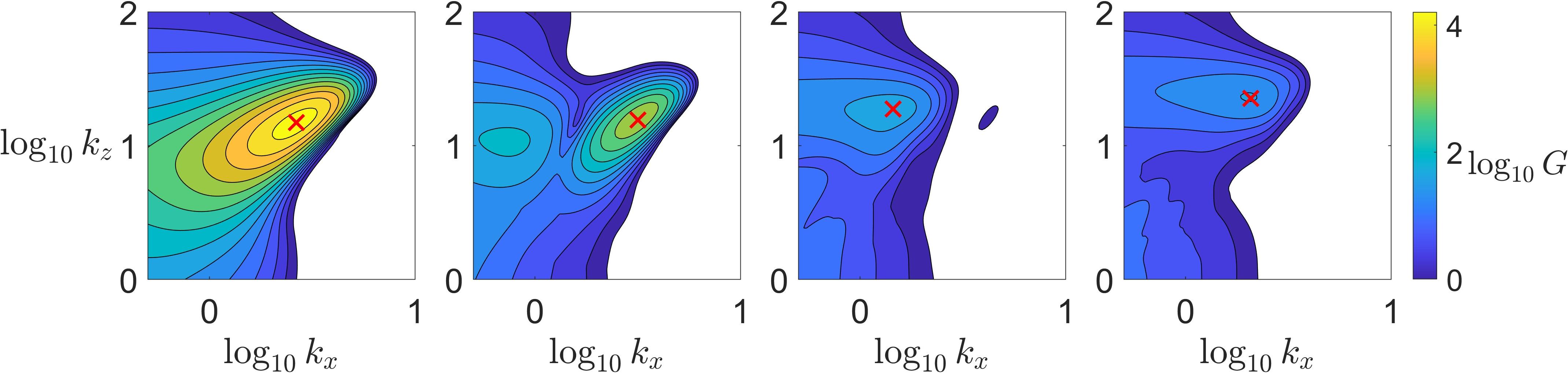}
        %\caption{}
    \end{subfigure}
\caption{Three sequences of contour plots of the optimal gain in wavenumber space, close to the optimal wavenumber pair identified in the unstratified flow, for parameters $T=5$, $\beta=1$, $\Rey=200$ and $\Pran=7$. The first row is for the minimal system, the second row for the reduced system, and the third row for the full system with $M=10$. The columns show different stratification strengths $N$ as indicated above each column, increasing from left to right. A red cross is used to indicate the global optimal growth point. This plot demonstrates that increasing $N$ to $N_c$ in each system significantly reduces the possible transient growth to the point where it is either removed completely or damped enough so that another mechanism becomes dominant.}
\label{WNseq}
\end{figure}
For non-trivial solutions
\renewcommand{\arraystretch}{1}
\begin{equation}
h_1^2 \sigma^4 + \left[\,(2k_x^2+k_z^2)N^2  + \beta^2 k_x^4 \,\right]\sigma^2 + \left[ \,k_x^2 N^4+ \tfrac{1}{2}\beta^2 k_x^2(k_z^2-2k_x^2)N^2 - \tfrac{1}{4}\beta^4 k_x^4(k_z^2-k_x^2)\,\right] = 0
\label{C5eig2}
\end{equation}
which is a quadratic in $\sigma^2$. As a result, if $\sigma$ is a solution, so is $-\sigma$ and then the only way to have stability is for $\sigma$ to be purely imaginary ($\sigma^2 <0$). Consequently, the threshold for instability is given by a root $\sigma^2=0$ which occurs when
\beq
\left[\,N^2- \tfrac{1}{2} \beta^2 k_x^2\, \right]\left[\,N^2- \tfrac{1}{2}\beta^2(k_x^2-k_z^2)\,\right] = 0
\eeq
indicating instability (the lhs is negative) for 
\begin{equation}
 -\tfrac{1}{2}\beta^2(k_z^2-k_x^2) \,<\, N^2 \,<\, \tfrac{1}{2}\beta^2 k_x^2.
\label{C5cond2b}
\end{equation}
For $N^2=0$, it is known \cite{OK25} that there is linear instability when $k_z > k_x$ and now we can see here that this is stabilised when $N \geq N_c:=\tfrac{1}{\sqrt{2}}\beta k_x$. 
In the opposite scenario, $k_x \geq k_z$, however, we also see a  new instability if the stratification is in the range 
\beq
0\,\leq \,\tfrac{1}{2} \beta^2 (k_x^2-k_z^2) \,<\, N^2 \,<\,  \tfrac{1}{2} \beta^2 k_x^2.
\eeq
This new stratified instability is not found to be important when growth is optimised over all wavenumber pairings $(k_x,k_z)$ in the full system and so is not pursued further here.

We now look at the effect of horizontal stratification on the energy growth optimised over the minimal, reduced and full systems. There are two key additions in doing this beyond including more spanwise wavenumbers: the vertical velocity and diffusion. Both restrict the growth to a finite time horizon. Diffusion also preferentially arrests the growth of the larger wavenumbers, leaving only a shrinking band of $k_x$ wavenumbers with energy growth as stratification is increased.  The largest such $k_x$ (for $N^2=0$) could be used to provide an upper bound for the critical stratification, $N_c$, which turns the growth off. But a better choice adopted below is to take the $k_x$ which gives the global optimal as it requires the most damping due to its superior growth. In fact, the last remaining growing $k_x$, as $N$ is increased, may well be somewhere in between the two options but this is not straightforward to predict.

\begin{figure}
    \centering
        \includegraphics[width=\textwidth]{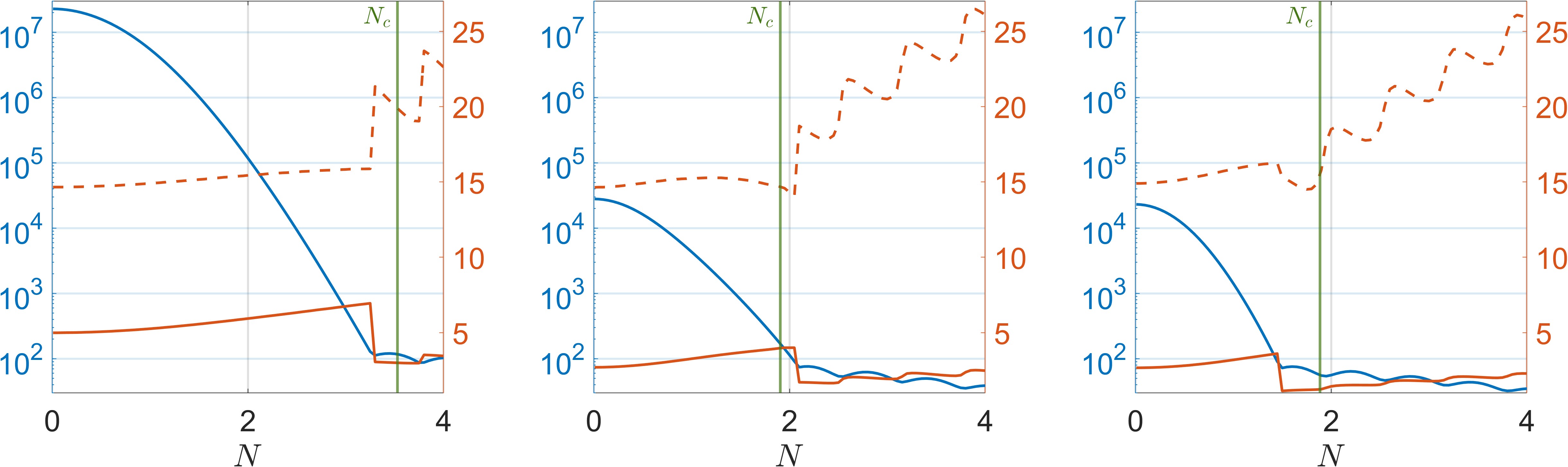}
\caption{Plots of the global optimal gain (blue) and associated optimal $k_x$ (solid red) and $k_z$ (dashed red) against the stratification strength $N$, for parameters $T=5$, $\beta=1$, $\Rey=200$ and $\Pran=7$. The first plot shows the minimal system, the second shows the reduced system, and the third shows the full system with $M=10$. A solid green vertical line shows the location of the critical stratification strength $N_c$ in each case, calculated using the $k_x$ value of the global optimal gain in the associated unstratified problem. This plot demonstrates both the significant reduction in transient growth with $N$, as well as the switch to a different mechanism when stratification has damped the growth enough. }
\label{c3vN}
\end{figure}

Figure \ref{WNseq} displays a sequence of four plots of the optimal gain in wavenumber space, close to the pair that maximises this optimal as the stratification increases for the minimal, reduced and full ($M=10$) systems. In all three, there is a general decrease in growth as $N$ increases, as vertical motions $w$ become more and more damped by stratification. The position of the global optimal also qualitatively changes at some point - for the minimal and reduced systems between  $N=2$ and $N=3.5$ and between $N=1$ and $N=2$ for the full system - indicating a change in the preferred transient growth mechanism. Figure \ref{c3vN} makes this clear by showing the discontinuous jumps in the optimal wavenumbers as $N$ varies and also indicates that the critical stratification $N_c$ derived from the minimal system does a reasonable job in predicting this change ($N_c\approx 3.53$, $\approx 1.91$ and $\approx 1.89$ for the minimal, reduced and full systems respectively). 

%
% fig 8
%

%The switch of the optimal gain away from the push-over-Orr symbiosis to a different mechanism, already observed in figure \ref{WNseq}, is made extremely clear using these new plots; a sharp change in the value of both wavenumbers coincides with a change of behaviour in the profile of gain against $N$. For all three models, the calculated value for $N_c$ (using \eq{C5iff} and the optimal $k_x$ value in the associated unstratified problem) provides an excellent prediction, which is indicated by the proximity of the vertical green lines with the sharp change in optimal wavenumbers. After the switch, the new mechanism is still damped mildly by increasing stratification, but the damping is not as strong and the gain no longer shows monotonic decay; in fact the gain seems to oscillate. This oscillatory behaviour of the gain as $N$ is varied, along with the varying of the optimal wavenumbers, is not pursued further here as the underlying mechanism is not push-over-Orr symbiosis, and the growth values themselves are not large. In fact, once the exponential growth is damped significantly so that it is no longer the optimal, there is no guarantee that the data here captures the fully global optimal, as the search over wavenumber space was restricted to the domain shown in all plots of figure \ref{WNseq}. At gain values this small, it is plausible that other mechanisms can produce larger (but same order of magnitude) growth, such as the traditional lift-up and Orr mechanisms (or in fact a combination of both). 

%
% fig 9
%
\begin{figure}
    \centering
        \includegraphics[width=0.4\textwidth]{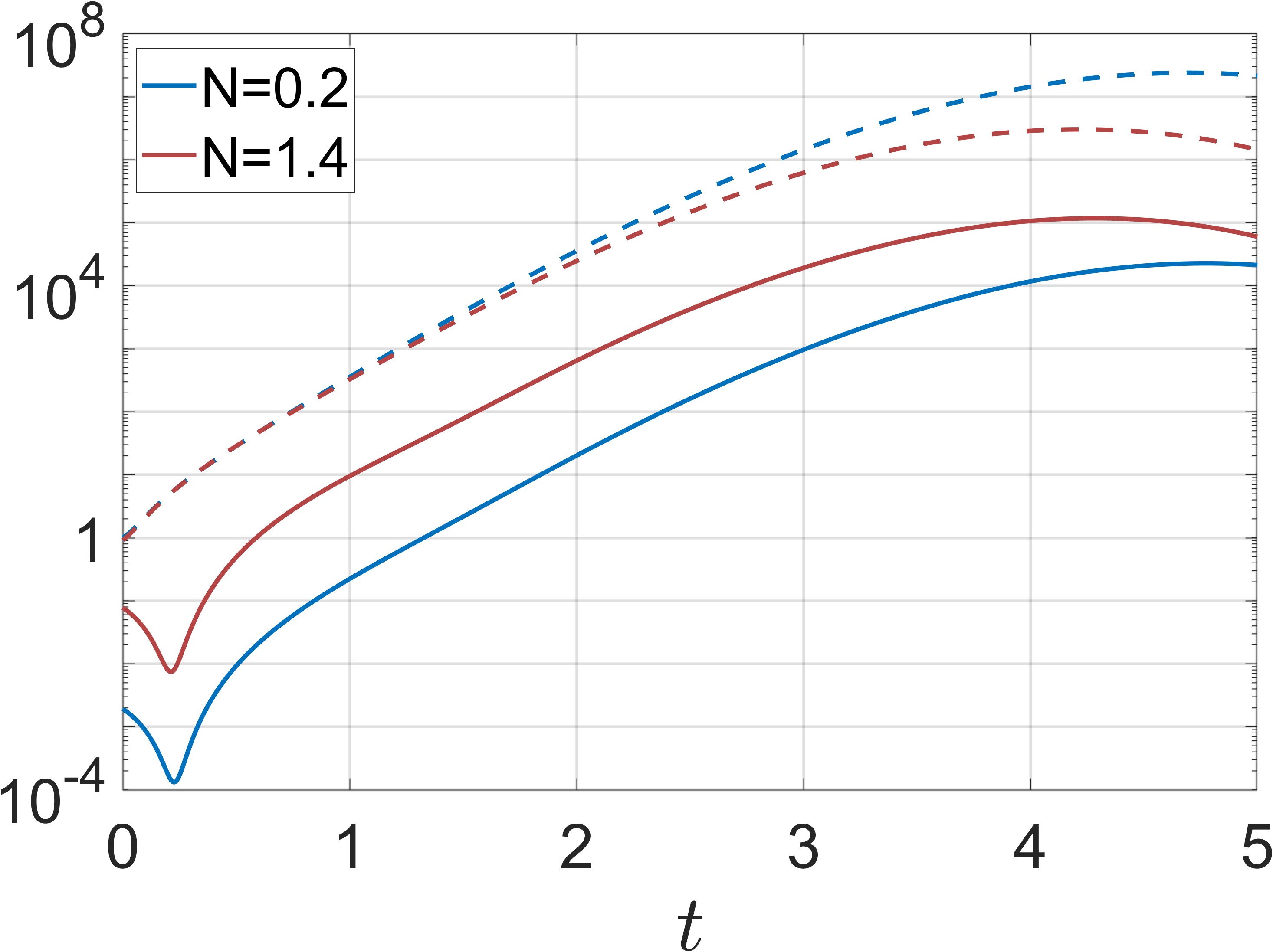}
    \hspace{1.5cm} % Adds space between subfigures
        \includegraphics[width=0.4\textwidth]{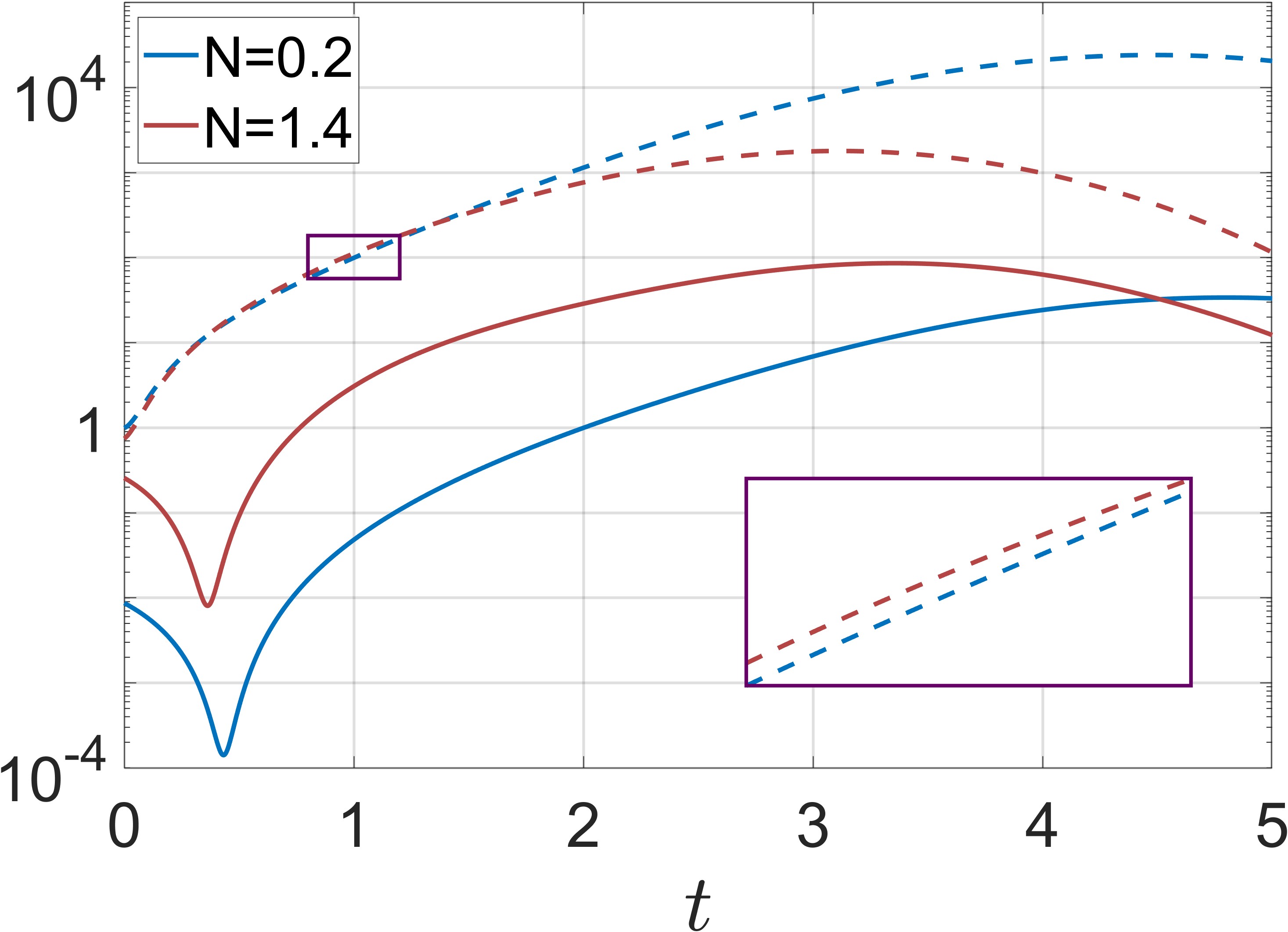}
    \caption{Left: energy breakdown in time for the global optimal perturbation in the minimal system, using parameters $T=5$, $\beta=1$, $\Rey=200$, $\Pran=7$ and two $N$ values. The dashed lines display the kinetic energy evolution while the solid lines display the potential energy evolution.  Right: same as the left figure, but for the the full system with $M=10$. The inset shows a zoom of the kinetic energy close to $t=1$, indicated by the purple rectangle. This figure shows that there are two phases in the growth of energy, but the potential energy always remains sub-dominant.}
\label{hshearenergy}
\end{figure}

Finally, a breakdown of the total energy into kinetic and potential energies as it evolves with time is shown in figure \ref{hshearenergy} for the  global optimal for two different stratifications and  both the minimal and full systems.
The potential energy growth in either system increases with larger stratification, although the potential energy remains sub-dominant to the kinetic energy at all times. Again there is some initial adjustment of the potential energy before it grows but the stratification here is much smaller than that used in the vertical shear case and there are no appreciable internal wave oscillations initially.

%---------------
%
% V Discussion
%
%---------------

\section{Discussion} \label{C5disc}

% what has been done

The motivation for this work was the recent (and related) observations that 
1. the lift-up mechanism  tends to inhibit the transient growth available on streaky flows in wall-bounded flows \cite{Loz-21,MK24,OK25}, 
and 
2. that artificially removing the wall-normal velocity can unlock considerably more growth \cite{OK25}. 
Using the simplified wall-less model of \cite{OK25}, we have confirmed here that introducing stable stratification which suppresses wall-normal velocities (the `vertical shear' case) can unlock essentially  all of the  enhanced growth seen in the unstratified minimal model of \cite{OK25} where  the wall-normal velocity was artificially set to zero. In contrast, imposing stable stratification such that the spanwise velocities are suppressed (the `horizontal shear' case) not unexpectedly inhibits growth by weakening the push over mechanism \cite{Loz-21,OK25}.  A formula for the critical stratification strength to suppress the growth mechanism is determined which proves a useful predictor for what is seen in the full numerical solutions.

% implications

The growth-enhancing effect of  stable stratification in the vertical shear case, however, comes at a cost for the complementary roll-to-streak process central to the self-sustaining near-wall cycle. This is thought entirely lift-up driven and so enhancing the streak-to-roll part of the cycle has correspondingly negative consequences for the ability of the rolls to regenerate the streaks. 
How this all plays out as the stratification changes is not completely clear as, on the one hand, larger growth on the  streaks  should be magnified quadratically through the nonlinearity,  while on the other, the feedback onto the rolls should be compromised by the suppression of wall-normal motions again. The latter was certainly seen in \cite{Eaves15} (see also \cite{Deguchi17, Olvera17}) where increasing stratification always eventually killed the exact coherent states existant for the unstratified flow although the situation can be more nuanced for different Prandtl numbers \cite{Langham20}.

% future work

In terms of future work, it would certainly be interesting to confirm the destabilizing effect of wall-normal stable stratification on streaky flows in targeted direct numerical simulations as performed in \cite{Loz-21} for  unstratified channel flow. There, the authors comment that removing the lift-up term is destabilizing as perturbations near the wall are enhanced (see p37 and figure 24(b) of \cite{Loz-21}) but this obviously falls well short of artificially removing all wall-normal velocities as done in \cite{OK25}. The key result here indicates that this extreme scenario is actually the best model of what stable stratification can do naturally which  does suggest that substantially  more transient growth could be available for stably-stratified wall-bounded streaky flows.

\bibliography{references}
\bibliographystyle{plainnat}

\end{document}